\documentstyle[epsf,floats,twocolumn,prl,aps]{revtex}
\newcommand{\BS}{\mbox{\boldmath $S$}}
\newcommand{\BN}{\mbox{\boldmath $n$}}
\newcommand{\BM}{\mbox{\boldmath $m$}}
\newcommand{\BL}{\mbox{\boldmath $l$}}
\begin{document}
\draft
\preprint{}
\title{Spin Chains with Periodic Array of Impurities}
\author{ Takahiro Fukui\cite{Email}} 
\address{Institute of Advanced Energy, Kyoto University,
Uji, Kyoto 611, Japan}
\author{Norio Kawakami}
\address{Department of Applied Physics,
Osaka University, Suita, Osaka 565, Japan}
\date{November 28, 1996: Revised February 27, 1997}
\maketitle
\begin{abstract}
We investigate  the spin chain model composed of periodic array of
two kinds of spins $S_1$ and $S_2$, which  
allows us to study the spin chains with impurities as well as 
the alternating spin chains in a unified fashion.
By using the Lieb-Shultz-Mattis theorem,
we first study the model rigorously,
and then by mapping it to the non-linear sigma model,
we extensively investigate low-energy properties 
with particular emphasis on the competition between the massive 
and massless phases. 
\end{abstract}
Quantum spin chains have been providing  a number of hot topics
not only in condensed matter physics but also in 
statistical physics, quantum field theory, etc.
In particular, the competition between the massive and
massless states in the spin chains has attracted 
much current interest. 
For example, doping impurities into massive systems such as 
the Haldane-gap system \cite{HalGap}, 
the two-leg ladder system\cite{RICE,NTAFT}
and the spin-Peierls system \cite{RRDR}
may cause the massless states, and sometimes stabilize
the long-range magnetic order. For these systems it is an
interesting issue to
clarify how the quantum coherence, which produces spin gaps, is
suppressed by impurities to result in a  gapless state\cite{Theor}.
Also, if the impurities are magnetic and their concentration
becomes high with periodic arrangement, the
system naturally leads to the alternating spin chains 
which have also been studied intensively \cite{ALTER}. 
A common feature in these problems is 
how the massive and massless states 
compete with each other, providing a variety 
of interesting phenomena.

In this paper we investigate  
the quantum spin chain  composed of 
the periodic array of two kinds of spin $S_1$ and $S_2$, 
putting particular emphasis on the formation 
of the massive and massless states. The model proposed here 
is related to  the interesting topics mentioned above, and 
naturally interpolates the impurity models 
and the alternating spin chain models.
For example, if we consider the case of dilute $S_1$ 
spins in the background of the host spins with integer $S_2$,
the model is  related to the  Haldane gap systems with 
 magnetic impurities.  
On the other hand, in the high 
density limit of $S_1$ spins,
it describes the alternating spin chains.
We will investigate low-energy properties of the model
by using the Lieb, Schultz and Mattis (LSM) theorem and  
non-linear sigma model techniques.

The model we will investigate 
consists of two kind of spins with nearest neighbor interaction. 
The Hamiltonian is given by
\begin{equation}
H=\sum_{j=1}^{N}J_j\BS_j\cdot\BS_{j+1},
\label{Ham1}
\end{equation}
with the periodic boundary condition $\BS_{j+N}=\BS_{j}$,
where $N~(=MN')$ is the number of sites, which is 
assumed to be even integral in what follows.
The spin quantum number $S_j$ on each site is
$S_j=S_1$ for $j=1$ (mod $M$) and $S_j=S_2$ for others:
The periodic array of $S_1$ impurities is embedded
in the host spins $S_2$ with  the period $M$, i.e,
\begin{equation}
\underbrace{S_1\otimes S_2\otimes S_2\otimes\cdots\otimes S_2}_{M}
\otimes S_1\otimes S_2\otimes\cdots\cdots\otimes S_2.
\end{equation}
Therefore by tuning the period $M$, we can naturally interpolate
the dilute impurity models (large $M$) and the alternating spin chains
($M$ being order of unity).
The antiferromagnetic coupling $J_j~(>0)$ is assumed to have
a bond-dependence, $J_j=J(1+\gamma)$ for $j=0$ and $1$ (mod $M$; 
between spins with $S=S_1$ and $S_2$) 
and $J_j=J$ for others (between the same spins with $S=S_2$).

Let us start by specifying the  ground state properties.
Since spins are on a bipartite lattice,
we can apply the Marshall theorem\cite{Aue} to (\ref{Ham1}):
The ground state is specified by the spin quantum number
$S=0~(|S_1-S_2|N/M)$ for $M=$ odd (even),
which is non-degenerate except for the trivial spin degeneracy.
Therefore the ground state of our model is 
either a spin singlet or ferrimagnetic. 
We are interested in the quantum effects on the spin liquid phase,
so that we will mainly concentrate on the singlet cases (odd $M$) which
possess a variety of interesting properties. 
We shall show the simple but remarkable fact  
that the universality class
of the model is solely determined 
by the impurity spin $S_1$, as far as the 
concentration of $S_1$ spins is finite.

To begin with, we apply the LSM theorem\cite{LSM} in order
to address a question whether the above model 
can have a gapless excitation.
Let $T^{M}$ be the $M$-sites translation operator,
which commutes with the Hamiltonian (\ref{Ham1}) by definition.
Therefore, together with the Marshall theorem, we find that
it acts on the ground state $|\Phi_0\rangle$ as
$T^{M}|\Phi_0\rangle=e^{i\phi}|\Phi_0\rangle$.
Next define the twist operator 
$U=\exp(\frac{2\pi i}{N}\sum_{j=1}^NjS_j^z)$,
and  the corresponding twisted state $|\Phi\rangle=U|\Phi_0\rangle$.
The energy increment due to twist is easily calculated as
$\langle\Psi|H|\Psi\rangle-\langle\Psi_0|H|\Psi_0
\rangle<{\rm const.}/N$.
In order to ensure that the twisted state $|\Phi\rangle$
is actually an excited state, we need to show
the orthogonality of these states. We immediately find
$T^MUT^{-M}=(-1)^{2S_{\rm eff}M}U$ with 
\begin{equation}
S_{\rm eff}=\left[S_1+(M-1)S_2\right]/M.
\label{Sef}
\end{equation}
Note that the orthogonality condition is satisfied by 
the minus sign of the factor $(-1)^{2S_{\rm eff}M}$.
Therefore, in the case $2S_{\rm eff}M=$ odd-integer,
we can prove that the twisted state 
is orthogonal to the ground state $\langle\Phi_0|\Phi\rangle=0$,
and hence the system has a gapless excitation,
or alternatively, degenerate ground states.
For example, this is the  case for the spin chain with
$S_1=1/2$ and $S_2=1$. In general, 
the existence of an excited state 
with  $O(1/N)$ energy depends on how many
half-integer spins are included in the unit composed of $M$ sites.

Here  we  deal with the  most interesting case, i.e.
the integer $S_2$ spin model with $S_1$ spins doped,
which may be related to the Haldane gap system with magnetic impurities. 
According to the above theorem, it is seen in this case:
it is the impurity spin $S_1$ that controls whether the 
system is massless (or massive with degenerate 
ground states) in the integer $S_2$ background. 
For dilute impurities with half-integer $S_1$, an excited 
state with energy of $O(1/N)$ 
ensured by the LSM theorem may be essentially 
the same as an impurity state in the Haldane gap systems.
According to extensive work on the Haldane-gap systems,
doping impurities induces the 
free $S=1/2$ degrees of freedom\cite{Ken} near the edge of the
valence-bond-solid states\cite{AKLT},
and they  couple with the doped impurity spin $S_1$, 
making a local object of the effective impurity state.
\cite{HalImp}
If we increase the concentration of $S_1$ spins, 
correlations among these local impurity states 
become strong, and impurity bands are naturally formed by coherent
motion of effective impurity spins.
Owing to the quantum effects, they again become massless 
(massive, see below) for half-integer (integer) $S_1$,
which will be discussed below in terms of field theoretic methods.

Based on the above rigorous results, we now  construct
low-energy effective field theory, which allows
us to  study the model with integer-$S_1$
 as well as half-integer-$S_1$ cases
in a unified way.  Moreover, we can see
which possibility for the ground state is realized for  
 half-integer-$S_1$ cases, i.e. a massless state or
a massive state with degenerate ground state.
For this purpose, we shall map the system (\ref{Ham1}) to 
the non-linear sigma model\cite{Hal,Rev}
by the use of the SU(2) coherent state path-integrals\cite{Coh}.
In what follows, we again concentrate ourselves 
on the systems with singlet
ground state, i.e, for the case $M=$ odd.
Note that the following analysis can be 
applied to the system with {\it finite concentration} of 
impurities in the thermodynamic limit.

The coherent state in the spin-$S$ representation is here defined by
$|\zeta\rangle=(1+|\zeta|^2)^{-S}\exp(\zeta S^-)|S,m=S\rangle$.
Parametrizing $\zeta=\tan\frac{\theta}{2}e^{i\phi}$,
we have $\langle\zeta|\BS|\zeta\rangle=S\BN$,
where $\BN=(\sin\theta\cos\phi,\sin\theta\sin\phi,\cos\theta)$.
By using the over-completeness relation 
$\int d\mu(\zeta)|\zeta\rangle\langle\zeta|=1$ with the invariant measure
$d\mu(\zeta)=\frac{2S+1}{\pi}\frac{d^2\zeta}{(1+|\zeta|^2)^2}$,
we can derive the path-integral representation of the partition function.
Namely, by staggering the spin configuration as 
$\BS_j=S_j(-)^{j+1}\BN(j)$, 
the partition function $Z={\rm tr}\exp(-\beta H)$ of the system
(\ref{Ham1}) can be represented by  
$Z=\int\prod_{j=1}^N{\cal D}\mu[\BN(j)]\exp(-S)$,
where the action is explicitly given by
\begin{eqnarray}
S=i\sum_j^N&&(-)^jS_j\omega[\BN(j)]
\nonumber\\
&&-\sum_{j=1}^NJ_jS_jS_{j+1}\int_0^\beta\!\! d\tau\BN(j)\cdot\BN(j+1).
\label{Act}
\end{eqnarray}
Here $\omega[\BN(j)]$ is the Berry phase acquired by the $j$th spin,
\begin{equation}
\omega[\BN(j)]=\int_0^1\!\!du\int_0^\beta\!\! d\tau
\BN(j)\cdot\left[\partial_\tau\BN(j)\times\partial_u\BN(j)\right].
\end{equation}
In what follows, we assume that 
the behavior of the field $\BN(j)$ is similar to that for
the uniform spin chain in the continuum limit.
Therefore, we can divide $\BN(j)$ into slowly varying part and
fluctuation around it as usual,
$\BN(j)=\BM(j)+a(-)^{j+1}\BL(j)$. 
This assumption should be confirmed to be 
valid by comparison with the results of the LSM theorem.

Let us first calculate the Berry phase term in the continuum limit,
\begin{eqnarray}
S_B&=&i\sum_{k=1}^M\sum_{j=1}^{N/(2M)}\left[
S_1\delta_{k,1}-S_2(-)^k(1-\delta_{k,1})\right]
\nonumber\\
&&\!\!\!\times
\left(\omega[\BN(2Mj-M+k)]-\omega[\BN(2Mj-2M+k)]\right)
\nonumber\\
&=&i\frac{S_1}{2}\int\!\!\!\int
d^2x\BM\cdot(\partial_1\BM\times\partial_2\BM)
\nonumber\\
&&\qquad
+iS_{\rm eff}\int\!\!\!\int d^2x\BL\cdot(\BM\times\partial_2\BM)
\end{eqnarray}      
with $S_{\rm eff}$ defined in eq.(\ref{Sef}),
where $x_1=x$ and $x_2=\tau$.
To derive this formula, we have taken the continuum 
limit with respect to $Ma$ lattice spacing:  
$\BN(2Mj-M+k)-\BN(2Mj-2M+k)\sim Ma\partial_1\BM(2Mj-M+k)
+(-)^k2a\BL(2Mj-M+k)$, where $a$ is the lattice constant.
This procedure implies that we are 
now concerned with the most important low-energy mode, although  
there are other massive excitation modes 
because the period of the lattice is $M$
in our model.  The validity of this procedure
will be discussed by comparing the 
results with those of the LSM theorem.
It should also be noted that the continuum limit here corresponds to
the limit $N/(2M)\rightarrow\infty$. Namely, we assume in this
approach the finite concentration $1/M$ of the $S_1$ spins.
Next, the interaction term is rewritten
in the continuum limit,
\begin{eqnarray}
S_I&=&-J\sum_{k=1}^{M}\sum_{j=1}^{N/M}
\bigl[(1+\gamma)S_1S_2(\delta_{k,1}+\delta_{k,M})
\nonumber\\
&&\qquad\qquad\qquad\qquad
+S_2^2(1-\delta_{k,1})(1-\delta_{k,M})\bigr]
\nonumber\\
&&\quad\times\int d\tau\BN(Mj-M+k)\cdot\BN(Mj-M+k+1)
\nonumber\\
&\sim&\frac{Ja}{2}\beta\int\!\!\!\int d^2x
\left[(\partial_1\BM)^2+4\BL^2\right]+{\rm const.}
\end{eqnarray}
with $\beta=S_2^2+2[(1+\gamma)S_1-S_2]S_2/M$\cite{Com1}.
In this way, we end up with the action composed of the fields 
$\BM(x)$ and $\BL(x)$.
The term  of $\BL(x)$ appears in a quadratic form, and can 
be easily integrated out, which consequently results in
the following Lagrangian density
\begin{eqnarray}
{\cal L}=\frac{1}{2g}&&\left[
v(\partial_1\BM)^2+\frac{1}{v}(\partial_2\BM)^2\right]
\nonumber\\
&&\qquad\qquad+\frac{\theta}{8\pi}\epsilon_{\mu\nu}\BM
\cdot(\partial_\mu\BM\times\partial_\nu\BM), 
\label{Lag}
\end{eqnarray}
with
\begin{eqnarray}
&&\theta=2\pi iS_1~(=2\pi iS_{\rm eff}M),
\nonumber\\
&&g=\frac{2}{S_2+(S_1-S_2)/M}~\left(=\frac{2}{S_{\rm eff}}\right),
\nonumber\\ 
&&v=2Ja\frac{S_2^2+2[(1+\gamma)S_1-S_2]S_2/M}{S_2+(S_1-S_2)/M},
\end{eqnarray}
where $S_{\rm eff}$ is defined in eq.(\ref{Sef}).

We recall here that the topological term with $\theta$ can 
completely specify whether the system is massive or massless.
By observing that only $S_1$ appears in 
$\theta$, we again arrive at the conclusion that 
{\it the universality class of the system is solely
determined by the impurity spin $S_1$},
which completely fits in with the results of the 
extended LSM theorem obtained above: for half-integer $S_1$, 
the system is massless even if $S_2=$ integer
because  $\theta=\pi i$ (mod $2\pi i$).
Furthermore, for integer $S_1$ we can say beyond the LSM theorem 
that the system should be massive because $\theta=0$ (mod $2\pi i$). 
By using the above formulae, we can discuss
the interesting results deduced by the LSM theorem in more detail.
The following statements are valid for the system with 
finite concentration of impurities.
(a) Haldane-gap systems ($S_2=$ integer) become massless 
(are still massive) for half-integer (integer) $S_1$ impurities.
(b) Massless spin chains ($S_2=$ half-integer) become massive
(are still massless) for integer (half-integer) $S_1$ impurities.
Especially in the latter case, we would like to recall the analysis by 
Eggert and Affleck\cite{EggAff}, from which 
we can naively expect the following scenario:
Integer $S_1$ impurities are screened by the two neighboring
half-integer spins $S_1$, forming local integer $S_2-2S_1$ objects.
These effective spins may couple with each other 
in a coherent way and produce a gap \`a la Haldane. 

We should mention  here that our mapping to the 
sigma model may be justified for a high or
intermediate concentration of 
impurities, but not for a dilute limit. However, a
qualitative feature whether the system is massive or not
is determined solely by the topological term,
and is expected to be correctly specified via the present analysis
so far as the impurity concentration is finite
in the thermodynamic limit.  

So far we have been mainly concerned with  the impurity 
effects on spin chains. 
Here we discuss the opposite limiting case
(high density of $S_1$ spins), i.e. the alternating spin chain
 system in more detail. 
For example, when $M=2$ with a half-integer $S_1$ and 
an integer $S_2$, our model (\ref{Ham1}) neatly describes the 
alternating spin chain system which has a 
ferrimagnetic ground state\cite{Ferri}, being consistent with 
experiments found so far\cite{ALTER}. 
In this connection, 
modified alternating spin chains with a singlet ground state 
(quantum liquid phase) have been actively investigated,
for which quantum fluctuations should play a vital role
\cite{VEGA,TONE}. 
This problem may provide a new interesting paradigm of spin 
chains bridging the massive and massless Heisenberg chains.
As a typical example\cite{TONE}, we here consider a slight extension 
of  the Hamiltonian (\ref{Ham1}) with 
$S_j=S_1$ for $j=1$ and 2 (mod 4) and $S_j=S_2$ for $j$=3 and 4 (mod 4),
namely,
\begin{equation}
S_1\otimes S_1\otimes S_2\otimes S_2\otimes S_1\otimes S_1\otimes
\cdots \otimes S_2\otimes S_2.
\label{Sys2}
\end{equation}
Antiferromagnetic couplings are 
$J_j=J(1-\gamma_1)$ for $j=1$ (mod 4; between $S_1$),
$J_j=J(1-\gamma_2)$ for $j=3$ (mod 4; between $S_2$)
and $J_j=J[1+(\gamma_1+\gamma_2)/2]$ for $j=$ others 
(between $S_1$ and $S_2$).  
According to the Marshall theorem, it is easily found that
the ground state of this model is singlet. 
Though the system (\ref{Sys2}) is invariant under 
four sites translation $T^4$, the LSM theorem cannot apply 
for any values of $S_1$ and $S_2$,
i.e, the twisted state is not orthogonal to the ground state.
This suggests that the system may be massive in general.
However, we should also note that since the system  
has spin-alternation as well as bond-alternation,
there could exist non-trivial massless phases between the 
massive phases, as is the case for 
uniform spin systems\cite{AffHal}.
We will clarify these points using non-linear sigma model techniques.

Keeping the above observations in mind, we derive the effective theory.
We have the same Lagrangian (\ref{Lag}), but with
\begin{eqnarray}
&&\theta=2\pi iS_{\rm eff}(1+\gamma_{\rm eff}),
\nonumber\\
&&g=\frac{2}{S_{\rm eff}\sqrt{1-\gamma_{\rm eff}^2}},
\nonumber\\
&&v=2Ja(S_{\rm eff}-\Delta S_\gamma)\sqrt{1-\gamma_{\rm eff}^2},
\label{TGV2}
\end{eqnarray}
where
\begin{eqnarray}
&&S_{\rm eff}=(S_1+S_2)/2,
\nonumber\\
&&\Delta S_\gamma=
(S_1-S_2)(\gamma_1S_1-\gamma_2S_2)/(4S_{\rm eff}),
\nonumber\\
&&\gamma_{\rm eff}=\frac{(S_1+S_2)(\gamma_1S_1+\gamma_2S_2)-(S_1-S_2)^2}
{(S_1+S_2)^2-(S_1-S_2)(\gamma_1S_1-\gamma_2S_2)}.
\label{Eff2}
\end{eqnarray}

We wish to study non-trivial cases, taking  $S_1=1/2$ and $S_2=1$ as
an example, and then generalize the discussions to arbitrary spin cases.
To begin with, let us set $\gamma_1=\gamma_2=0$. 
We find from eqs.(\ref{TGV2}) and (\ref{Eff2}) that
the topological term $\theta$ is different 
from $\pi i$ (mod $2\pi i$), and therefore
the system is  massive. 
Even in this case, $\gamma_{\rm eff}\ne0$ since it includes the
effects of the spin-alternation.
Let us next introduce the bond-alternation term $\gamma$, and observe 
what happens if the parameter $\gamma$  changes from
$-1$ to 1 for the case $\gamma_1=\gamma_2\equiv\gamma$. 
At $\gamma=-1$, the model (\ref{Sys2}) becomes a set of
isolated dimers, as seen from (\ref{Sys2}).
This system  has massive excitations, which 
is  indeed consistent with $\theta=0$ and
$g\rightarrow\infty$ in eq.(\ref{TGV2}).
If we increase $\gamma$ up to 1, 
the effective  $\gamma_{\rm eff}$ 
changes from $\gamma_{\rm eff}=-1$ to 1.
The corresponding value of $\theta$ changes from 0 to 
$\theta =3\pi i$. 
So it is predicted  that we encounter a massless fixed point 
once at $\theta =\pi i$ during this process. 
For the case $\gamma \rightarrow 1$, we need more 
careful treatment.
As naively expected, in this case, $S_1$ and $S_2$ spins strongly
couple with each other antiferromagnetically and 
the model (\ref{Sys2}) should behave 
like the uniform $S=1/2$ spin chain with massless excitations
(indeed we have $\theta\rightarrow  3\pi i$).
We should note, however, 
that exactly at $\gamma=1$ the system is separated into isolated 
pairs of spins, which may cause a singular behavior in 
the coupling constant, $g\rightarrow\infty$.
Therefore, for $\gamma\rightarrow1$ 
the model  exhibits a behavior quite  similar to that for
a half-integral  spin chain although it is still massive.

In general, for the case with  half-odd integer
$S_1$  and integer $S_2$, we can predict that there appear 
massless phases $2S_{\rm eff}-1/2$ times
as the bond alternation parameter 
changes from $\gamma=-1$ to 1.
It is also shown that for the cases where $S_1$ and $S_2$ are 
the same type of spins, i.e, half-odd
integers or integers, there appear $2S_{\rm eff}$ 
times massless fixed points.  We wish to 
note that in  the uniform case $S_1=S_2=S_{\rm eff}$
our formulae reduce to those for the ordinary spin 
chain with bond-alternation, which has
$2S_{\rm eff}$  critical points for $-1<\gamma<1$  \cite{AffHal}.
In this way, our approach based on non-linear sigma model techniques 
are quite powerful to systematically 
study the low-energy properties of alternating spin 
chain systems.  


The authors would like to thank M. Chiba
for valuable discussions.
This work is partly
supported by the Grant-in-Aid from the Ministry of
Education, Science and Culture, Japan.



\begin{references}
\bibitem{Com1} This $\beta$ is valid only for $M\ge3$.
If we include $M=1$, we should replace it to 
$\beta=S_1^2\delta_{M,1}
+\{S_2^2+2[(1+\gamma)S_1-S_2]S_2/M\}(1-\delta_{M,1})$.
\end{references}
\end{document}